# Highly efficient on-chip erbium-ytterbium co-doped lithium niobate waveguide amplifiers

YUQI ZHANG,[1] QIANG LUO,[1] DAHUAI ZHENG,[1,3] SHUOLIN WANG,[2] SHIGUO LIU,[1] HONGDE LIU,[1,*] FANG BO,[1,4] YONGFA KONG,[1,5] AND JINGJUN XU[1,6]

[1]*MOE Key Laboratory of Weak-Light Nonlinear Photonics, TEDA Institute of Applied Physics and School of Physics, Nankai University, Tianjin 300457, China*
[2]*School of Science, Jiangsu University of Science and Technology, Zhenjiang 212100, China*
[3]*dhzheng@nankai.edu.cn*
[4]*bofang@nankai.edu.cn*
[5]*kongyf@nankai.edu.cn*
[6]*jjxu@nankai.edu.cn*
*\*liuhd97@nankai.edu.cn*

**Abstract:** The ability to amplify optical signals is of paramount importance in photonic integrated circuits (PICs). Recently, lithium niobate on insulator (LNOI) has attracted increasing interests as an emerging PIC platform. However, the shortage of active devices on LNOI platform limits the development of optical amplification. Here, we firstly report an efficient waveguide amplifier based on erbium and ytterbium co-doped LNOI by using electron beam lithography and inductively coupled plasma reactive ion etching process. We have demonstrated that the net internal gain in the communication band is 15.70 dB/cm under the pumping of 974 nm continuous laser. Benefiting from the efficient pumping facilitated by energy transfer between ytterbium and erbium ions, signal amplification can be achieved at a low pump power of 0.1 mW. It is currently the most efficient waveguide amplifier under unidirectional pumping reported on the LNOI platform, with an internal conversion efficiency of 10%. This work proposes a new and efficient active device for LNOI integrated optical systems, which may become an important fundamental component of future lithium niobate photonic integration platforms.

## 1. Introduction

The ability to amplify optical signals holds vital importance in the fields of science and technology. Typically, signal amplification is achieved using rare-earth doped optical fibers or gain media based on III-V semiconductors [1]. Erbium (Er) doped fiber amplifier exhibits characteristics such as low nonlinearity and low noise amplification, which can cover the wide gain of C-band and L-band of telecommunications. It is the basis of the current long-distance optical fiber optical wave system. Compared with III-V semiconductors, rare-earth ion doped materials have the advantages of longer excited state lifetimes and less refractive index changes caused by doped ion excitation, which promotes the in-depth research on the PICs of $Er^{3+}$ doped waveguide amplifiers and lasers with various passive components on a chip-scale platform [2].

Lithium niobate ($LiNbO_3$, LN) is considered as one of the most promising photonic materials due to its excellent electro-, nonlinear- and acousto-optic properties, as well as its wide transparent window and relatively high refractive index [3,4]. Lithium niobate on insulator (LNOI) has strong optical constraints and retains good bulk characteristics, making it the main thin film platform for building chip integrated devices [5–8]. In the past few years, significant advancements have been made in the development of frequency comb sources [9,10], electro-optic modulators [11,12], frequency converters [13] and photodetectors [14] based on the LNOI platform. However, due to the indirect band gap structure of LN, it is difficult to achieve active optical gain, an important function in photonic integrated circuits. A promising solution for developing active devices based on LNOI is to doping rare earth ions into LN. In recent years, significant progress has been made in the integration of active components such as amplifiers [15–19] and lasers [20–26] based on the $Er^{3+}$ doped LNOI platform, demonstrating the enormous potential for achieving high-performance scalable light sources on the LNOI platform.

Considering the relatively weak pump absorption caused by the small absorption cross-section of $Er^{3+}$ at 980 nm, co-doping with $Yb^{3+}$ with a larger absorption cross-section can effectively enhance the pump absorption at 980 nm [2,27]. When $Yb^{3+}$ combine with $Er^{3+}$ in the same host, the excited $Yb^{3+}$ can transfer their energy to adjacent Er ions improving the pump efficiency [28–30]. Yet to date, $Er^{3+}/Yb^{3+}$ co-doped amplifier has not been demonstrated on the LNOI platform.

In this paper, the $Er^{3+}/Yb^{3+}$ co-doped LNOI waveguide amplifiers are firstly fabricated by electron beam lithography (EBL) and inductively coupled plasma reactive ion etching (ICP-RIE) processes. A 5 mm $Er^{3+}/Yb^{3+}$ co-doped waveguide amplifier has a net internal gain of 15.70 dB/cm at a signal wavelength of 1531 nm, and a power conversion efficiency of up to 10% between 980 nm and 1550 nm, representing the highest value with unidirectional pumping among $Er^{3+}$ doped LNOI waveguide amplifiers to date [16]. Furthermore, by analyzing the material properties of $Er^{3+}/Yb^{3+}$ co-doped, $Er^{3+}$ doped and $Yb^{3+}$ doped LN crystals, the effect and advantages of co-doping are clarified.

## 2. Fabrication

Fabrication of the $Er^{3+}/Yb^{3+}$ co-doped LNOI waveguide amplifiers starts from a Z-cut LNOI wafer which was ion-sliced from an $Er^{3+}/Yb^{3+}$ co-doped LN crystal grown by the Czochralski method. The preparation process is mainly divided into six steps, as shown in Fig. 1. Firstly, a LN boule with a doping concentration of 0.6 mol.% Er and 1.0 mol.% Yb was cut into wafers. Secondly, LNOI wafers were prepared using the "smart-cut" technology in cooperation with NANOLN. The thickness of the $Er^{3+}/Yb^{3+}$ co-doped LN (LN:Er,Yb), silicon-dioxide buffer layer, and silicon substrate are 0.6, 2.0 and 500 μm, respectively. Then a layer of hydrogen silsesquioxane (HSQ) in thickness was spin-coated on the wafer. Subsequently, the waveguide amplifiers mask was patterned by electron-beam lithography (EBL). Next, ICP-RIE machine with argon plasma etching was carried out to transfer the mask patterns into the $Er^{3+}/Yb^{3+}$ co-doped LN film, resulting in ridge waveguides with a 400 nm etching depth and a 60° wedge angle. Finally, the chip was immersed in buffered oxide etchant (BOE) solution for 5 min to remove the residual resist mask. Last, the chip facets were cleaved mechanically to ensure efficient fiber-to-chip coupling.

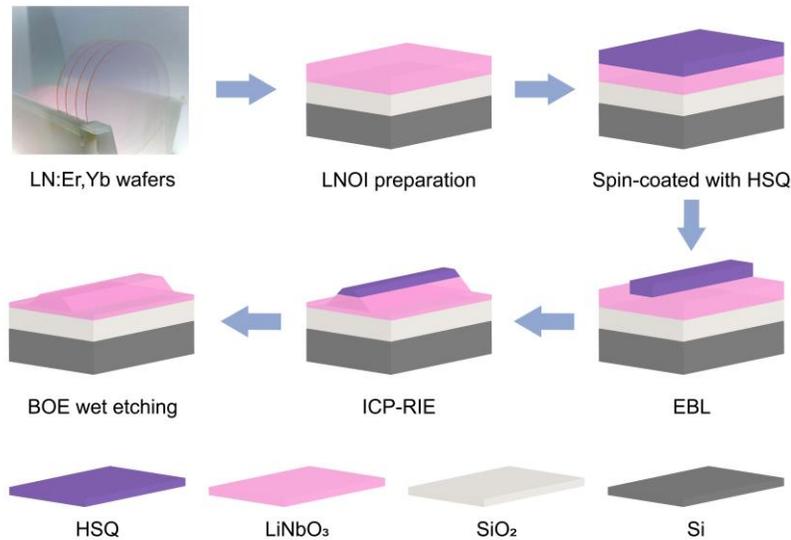

Fig. 1. Schematic of the fabrication process for $Er^{3+}/Yb^{3+}$ co-doped LNOI waveguides.

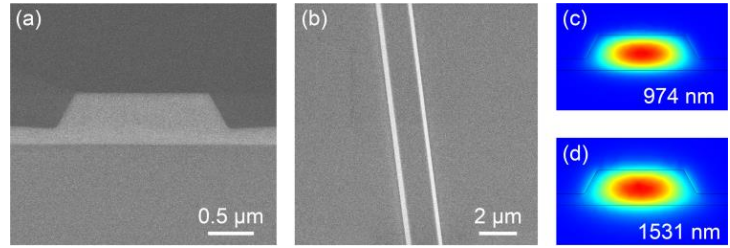

Fig. 2. SEM images of the (a) cross section and (b) longitudinal section of $Er^{3+}/Yb^{3+}$ co-doped LN waveguide. The simulated electric field distribution of single mode in the LN waveguide at (c) $\lambda = 974$ nm and (d) $\lambda = 1531$ nm.

Figures 2(a) and 2(b) present the scanning electron microscope (SEM) image of the cross section and longitudinal section of the fabricated $Er^{3+}/Yb^{3+}$ co-doped LN waveguide. The final length and top width of the straight waveguide is about 5 mm and 1.5 μm, respectively. Based on the structural parameters of the waveguide, we calculated the mode distribution of the basic transverse electrical modes at pump (~974 nm) and signal (~1531 nm) wavelengths, as shown in Figs. 2(c) and 2(d). The micron level ridge shaped LNOI waveguide has high refractive index contrast, resulting in light field limitations in both the 980 nm and 1550 nm bands.

### 3. Characterization

The experimental setup shown in Fig. 3 was used to illustrates the characterization of the LNOI amplifier. A continuous laser at 974 nm was selected as the pump to investigate the optical amplification performance of the $Er^{3+}/Yb^{3+}$ co-doped LNOI waveguides. A continuous-wave C-band tunable laser operating in the 1550 nm band was used as the signal. Before the pump and signal were launched into the chip, a variable optical attenuator (VOA), an optical coupler (OC) and a polarization controller (PC) were connected into the optical path to adjust the pump and signal power, split the light into two paths and optimize the polarization state, respectively. The pump and signal lights were combined by a wavelength division multiplexer (WDM) and launched into the $Er^{3+}/Yb^{3+}$ co-doped LNOI waveguide via a lensed fiber. Then, the amplified signal from the chip output facet was collected through a lensed fiber and emitted into an optical spectral analyzer (OSA) to detect the optical amplification performance of the LNOI waveguide. Simultaneously, the pump/signal power sent from the second port of OC were monitored by a power meter (PM).

Before characterizing the net internal gain of the $Er^{3+}/Yb^{3+}$ co-doped LN amplifier, we need to use whispering-gallery-resonator-loss measurements to characterize the pump light and signal light propagation losses of the waveguide. An $Er^{3+}/Yb^{3+}$ co-doped LN microring resonator with the radius of 100 μm was fabricated on the same chip using the same waveguide parameters. As shown in Fig. 4, the whispering-gallery modes at the resonance wavelength of 974 nm and 1531 nm were chosen for the measurement of the loaded $Q$ factor by fitting the transmission curves with a Lorentz function, giving $2.03\times10^5$ and $1.43\times10^5$ loaded $Q_L$, respectively. According to the gap between the microring and the waveguide, it was inferred that the 980 nm band was under-coupled and the 1550 nm band was over-coupled, so the intrinsic quality factors $Q_i$ were $4.60\times10^5$ and $2.24\times10^5$, respectively. Meanwhile, the effective refractive $n_{eff}$ at the relevant wavelength was calculated by $n_{eff}=\lambda^2/(2\pi R \cdot FSR)$, where $R$ and $FSR$ are the microring radius and free spectral range. Consequently, the propagation loss coefficient $\alpha$ was estimated using the equation:

$$\alpha = \frac{2\pi n_{eff}}{\lambda Q_i}, \quad (1)$$

where the $Er^{3+}/Yb^{3+}$ co-doped LN waveguide were deduced to be 0.73 dB/cm at 974 nm and 0.95 dB/cm at 1531 nm. It is worth noting that the propagation loss is mainly composed of waveguide scattering loss caused by sidewall roughness and absorption loss of erbium and ytterbium ions. Besides, considering the chip propagation loss and fiber-to-fiber insertion loss, we estimate that the fiber-to-chip coupling losses were 9.73 dB and 8.66 dB per facet at the wavelengths of 974 and 1531 nm. The high coupling loss is caused by the unoptimized mode field distribution in LN waveguide and lensed fibers. Based on the above calibration results, this paper refers to the pump power and signal power as on-chip power.

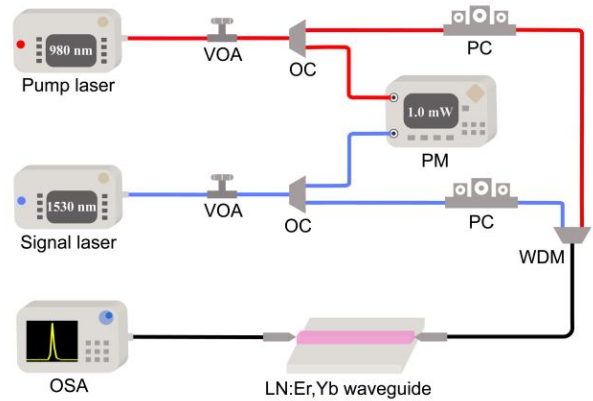

Fig. 3. Schematics of the experimental setup for gain characterization in $Er^{3+}/Yb^{3+}$ co-doped LNOI waveguide amplifiers.

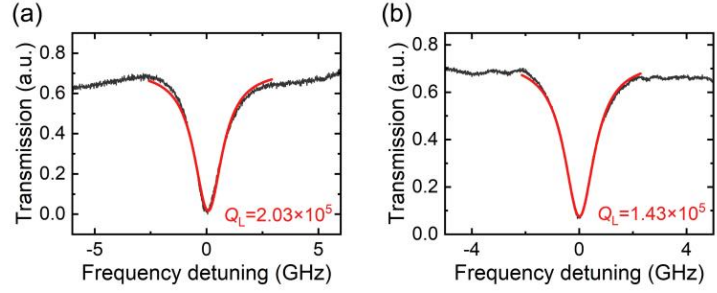

Fig. 4. Optical transmission spectra of $Er^{3+}/Yb^{3+}$ co-doped LNOI microring resonators on the same chip in (a) the 980 nm band and (b) the 1550 nm band. The Lorentz fit (red line) showing $2.03\times10^5$ and $1.43\times10^5$ loaded quality factors near 974 nm and 1531 nm, respectively.

The net internal gain of $Er^{3+}/Yb^{3+}$ co-doped LN waveguide amplifier was obtained by

$$g = 10\log_{10}\frac{P_{on}}{P_{off}} - \alpha L, \quad (2)$$

in which $P_{on}$ and $P_{off}$ denote the output signal power in pump-on and pump-off cases with coupling loss deducted, and $\alpha$ and $L$ respectively represent the propagation loss at the signal wavelength and the waveguide length. Therefore, $\alpha L$ is estimated as 0.48 dB.

Figure 5(a) shows the evolution of the measured signal spectrum with an increase of pump power at a fixed signal power. The dependence of net internal gain on pump power of the integrated amplifier with fixed signal power ∼28 nW at 1531.3 nm is shown in Fig. 5(b). It can be observed that the signal begins to be amplified at a lower pump power (~0.1 mW). The initial net internal gain rapidly increases with pump power. Subsequently, when the pump power is greater than 1 mW, the net gain tends to saturate. Continue to increase the pump power, and the net internal gain remains stable. A maximum net internal gain of 6.46 dB is achieved at a pump power of 6.20 mW, corresponding to a net gain per unit length of 12.92 dB/cm. Compared to $Er^{3+}$ doped LN straight waveguide amplifiers of the same length,

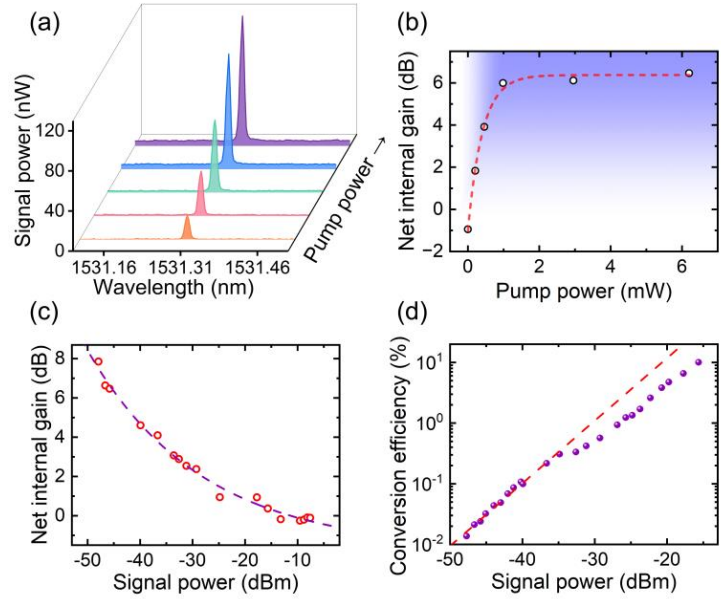

Fig. 5. (a) Measured signal spectra at ∼1531.31 nm with increasing pump powers of 0, 0.21, 0.46, 2.96 and 6.20 mW. (b) Dependence of net internal gain on pump power at a fixed on-chip input signal power of ~28 nW. (c) Net internal gain as a function of increasing signal power at a fixed pump power of ~6.20 mW. (d) The measured internal conversion efficiency (purple dot) is used as a function of signal power. The red dashed line shows a linear trend based on small signal gain.

the net internal gain is significantly improved at a low signal optical power.

In addition, we characterized the gain dependence of waveguide amplifier on the signal power, as shown in Fig. 5(c). The fixed pump power is 6.20 mW, and the signal power is adjusted from -48 dBm to -7 dBm. When the signal power is approximately -48 dBm, the maximum net internal gain is ~7.85 dB (15.70 dB/cm). Notably, with the increase of signal power, gain saturation can be observed, which is due to the depletion of the excited state population of erbium ions. On this basis, we further estimated the internal conversion efficiency $\eta$ of the $Er^{3+}/Yb^{3+}$ co-doped LN waveguide amplifier by following equation [16]

$$\eta = \frac{P_{on} - P_{off}}{P_{pump}} \times 100\%, \tag{3}$$

where $P_{pump}$ is on-chip input power with coupling loss deducted. As depicted in Fig. 6(d), the measured conversion efficiency increases linearly with the signal power in the small signal regime, then gradually deviates from the linear trend as the signal optical power increases. At a maximum signal power of -15 dBm, the internal conversion efficiency reached 10%. The internal conversion efficiency is currently the highest reported value for on-chip $Er^{3+}$ doped LNOI waveguide amplifiers. This result may be related to the co-doping of $Yb^{3+}$, which improves pump efficiency through energy transfer.

In order to study the superiority of 0.6 mol.% $Er^{3+}$ and 1.0 mol.% $Yb^{3+}$ co-doped LN (LN:Er,Yb), we grew 0.6 mol.% $Er^{3+}$ doped LN (LN:Er) and 1.0 mol.% $Yb^{3+}$ doped LN (LN:Yb) by the Czochralski method for exploration. Figure 6(a) shows the infrared absorption spectra of LN:Er,Yb, LN:Er and LN:Yb. It can be clearly seen that the absorption band of LN:Er,Yb is the superposition of the absorption bands of LN:Er and LN:Yb, which can be excited over a wider range and has higher absorption in the 980 nm band, resulting in higher pump efficiency. Figure 6(b) shows the infrared emission spectra of LN:Er,Yb, LN:Er and LN:Yb under 980 nm excitation at high pump power. It can be observed that the peaks of LN:Er and LN:Yb are at 1531 nm and 1062 nm, respectively. At this point, the peaks of LN:Er,Yb are close to LN:Er in the 1550 nm band. However, at lower pump power, the peaks of LN:Er,Yb are much higher as shown in the illustration. It indicates that the co-doping of $Yb^{3+}$ enormously improves the pump efficiency.

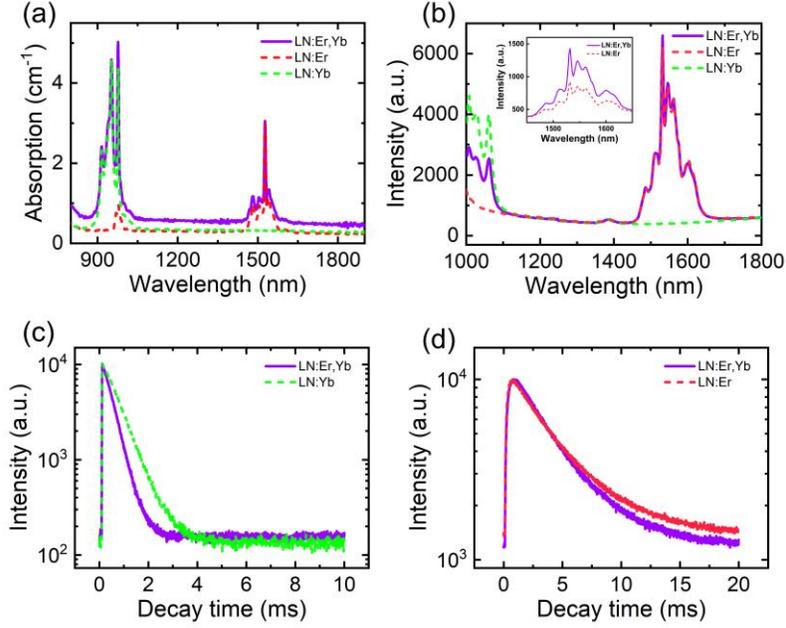

Fig. 6. (a) Infrared absorption spectra of $Er^{3+}/Yb^{3+}$ co-doped, $Er^{3+}$ doped and $Yb^{3+}$ doped LN. (b) Infrared emission spectra of $Er^{3+}/Yb^{3+}$ co-doped, $Er^{3+}$ doped and $Yb^{3+}$ doped LN under 980 nm excitation at high pump power. (The illustration is the infrared emission spectra of $Er^{3+}/Yb^{3+}$ co-doped and $Er^{3+}$ doped LN at lower pump power.) (c) Decay curves of the $Yb^{3+}$ emission at 1062 nm in $Er^{3+}/Yb^{3+}$ co-doped and $Yb^{3+}$ doped LN, excited with 980 nm. (d) Decay curves of the $Er^{3+}$ emission at 1531 nm in $Er^{3+}/Yb^{3+}$ co-doped and $Er^{3+}$ doped LN, excited with 980 nm.

The decay behavior of LN:Er,Yb and LN:Yb transitions measured at $Yb^{3+}$ ($\lambda_{ex}$ =980 nm) at 1062 nm is shown in Fig. 6(c). All attenuation curves can be well fitted by a function of time $I(t)=Ae^{(-t/\tau)}$, where A denotes the fitting amplitude and $\tau$ denotes the decay time. The emission decay times of $Yb^{3+}$ in LN:Er,Yb and LN:Yb are 0.40 ms and 0.64 ms, respectively. Energy transfer leads to faster decay of $Er^{3+}/Yb^{3+}$ co-doped LN, that is the energy transfer to $Er^{3+}$ constitutes an additional decay channel of $Yb^{3+}$ in its excited state. Based on this, we estimated the energy transfer efficiency from $Yb^{3+}$ to $Er^{3+}$ by $E=1-\tau_{Er,Yb}/\tau_{Yb}$ [31], here $\tau_{Er,Yb}$ and $\tau_{Yb}$ are the emission decay times of $Yb^{3+}$ in LN:Er,Yb and LN:Yb. The energy transfer efficiency is approximately 37.5%, which makes the pumping in the 980 nm band more efficient. The decay behavior of LN:Er,Yb and LN:Er transitions measured at $Er^{3+}$ ($\lambda_{ex}$=980 nm) at 1531 nm is shown in Fig. 6(d). $Er^{3+}$ has emission decay times of 3.52 ms and 3.77 ms in LN:Er,Yb and LN:Er, respectively. This indicates that $Er^{3+}/Yb^{3+}$ co-doping hardly affects the lifetime of Er.

## 4. Conclusions

In summary, we manufactured on-chip $Er^{3+}/Yb^{3+}$ co-doped LNOI waveguide amplifiers for the first-time using EBL and ICP-RIE processes. Under pumping in the 980 nm band, the communication band amplifier achieved a maximum net internal gain of approximately 15.70 dB/cm in a 5 mm long chip. Under a low pump power of 1 mW, the net internal gain of the amplifier approaches saturation. Compared with $Er^{3+}$ mono-doped amplifiers, the internal conversion efficiency is greatly improved. By analyzing the material properties, it was confirmed that the energy transfer from $Yb^{3+}$ to $Er^{3+}$ in LN:Er,Yb resulted in higher pump efficiency in the 980 nm band. In addition, by further optimizing the co-doping concentration of Er and Yb ions to achieve higher gain, growing crystals with better optical uniformity to reduce material losses, designing more suitable waveguide geometry to allow for high-power amplification, it is expected that the $Er^{3+}/Yb^{3+}$ co-doped LNOI waveguide amplifier will exhibit better amplification performance.

**Funding.** National Key Research and Development Program of China (Grant No. 2019YFA0705000); The National Natural Science Foundation of China (Nos. 12034010, 12134007); Natural Science Foundation of Tianjin (Grant Nos. 21JCZDJC00300, 21JCQNJC00250); Program for Changjiang Scholars and Innovative Research Team in University (No. IRT_13R29).

**Disclosures.** The authors declare no conflicts of interest.

**Data availability.** Data underlying the results presented in this paper are not publicly available at this time but may be obtained from the authors upon reasonable request.